# A new benchmark problem for electromagnetic modelling of superconductors: the high-$T_c$ superconducting dynamo

Mark Ainslie[1], Francesco Grilli[2], Loïc Quéval[3], Enric Pardo[4], Fernando Perez-Mendez[1], Ratu Mataira[5], Antonio Morandi[6], Asef Ghabeli[4], Chris Bumby[5], Roberto Brambilla[7]

[1]Department of Engineering, University of Cambridge, United Kingdom
[2]Institute for Technical Physics, Karlsruhe Institute of Technology, Karlsruhe, Germany
[3]Group of Electrical Engineering Paris (GeePs), CentraleSupélec, University of Paris-Saclay, France
[4]Institute of Electrical Engineering, Slovak Academy of Sciences, Slovakia
[5]Robinson Research Institute, Victoria University of Wellington, New Zealand
[6]University of Bologna, Bologna, Italy
[7]Retired; Ricerca sul Sistema Elettrico, Milano, Italy (formerly)

Corresponding author: mark.ainslie@eng.cam.ac.uk

Abstract

The high-$T_c$ superconducting (HTS) dynamo is a promising device that can inject large DC supercurrents into a closed superconducting circuit. This is particularly attractive to energise HTS coils in NMR/MRI magnets and superconducting rotating machines without the need for connection to a power supply via current leads. It is only very recently that quantitatively accurate, predictive models have been developed which are capable of analysing HTS dynamos and explain their underlying physical mechanism. In this work, we propose to use the HTS dynamo as a new benchmark problem for the HTS modelling community. The benchmark geometry consists of a permanent magnet (PM) rotating past a stationary HTS coated-conductor wire in the open-circuit configuration, assuming for simplicity the 2D (infinitely long) case. Despite this geometric simplicity the solution is complex, comprising time-varying spatially-inhomogeneous currents and fields throughout the superconducting volume. In this work, this benchmark problem has been implemented using several different methods, including **H**-formulation-based methods, coupled **H**-**A** and **T**-**A** formulations, the Minimum Electromagnetic Entropy Production method, and integral equation and volume integral equation-based equivalent circuit methods. Each of these approaches show excellent qualitative and quantitative agreement for the open-circuit equivalent instantaneous voltage and the cumulative time-averaged equivalent voltage, as well as the current density and electric field distributions within the HTS wire at key positions during the magnet transit. Finally, a critical analysis and comparison of each of the modelling frameworks is presented, based on the following key metrics: number of mesh elements in the HTS wire, total number of mesh elements in the model, number of degrees of freedom (DOFs), tolerance settings and the approximate time taken per cycle for each model. This benchmark and the results contained herein provide researchers with a suitable framework to validate, compare and optimise their own methods for modelling the HTS dynamo.

1. Introduction

The high-$T_c$ superconducting (HTS) dynamo [1-3] is a promising device that can inject large DC supercurrents into a closed superconducting circuit. It could be used, for example, to energise HTS coils in NMR/MRI magnets and superconducting rotating machines without the need for connection to a power supply via current leads [4, 5]. Despite the extensive experimental work carried out to date, comprehensively understanding the underlying physical mechanism of such dynamo-type flux pumps has proved challenging. A number of different explanations have been proposed to explain this mechanism [6-13], but quantitatively accurate, predictive calculations have been difficult. It was shown recently in Mataira *et al*. [14, 15] that the behaviour of the HTS dynamo can be explained well – most importantly, with good quantitative agreement – using classical electromagnetic theory. The DC output voltage obtained from an HTS dynamo arises naturally from a local rectification effect caused by overcritical eddy currents flowing within the HTS wire [6-8, 14, 16]: a classical effect that has been observed in HTS materials as far back as Vysotsky *et al*. [17]. The gap dependence of the open-circuit voltage computed by Ghabeli and Pardo [18] also agrees with experiments. In [18], it is also shown that this voltage is independent of the critical current density, $J_c$, when the superconductor is fully penetrated by supercurrents. Since these overcritical eddy currents must recirculate within the HTS wire, and can co-exist with a transport current, the wire width is a key parameter and [19] shows that this should be sufficiently large so that the eddy and transport currents do not drive the full width of the stator into the flux-flow regime.

A number of different numerical models have now been developed to simulate the electromagnetic behaviour of HTS materials. Such models represent useful and cost-efficient tools that provide insight into experimental results, as well as enable the optimisation and improvement of future device designs. To adequately compare the performance of different modelling approaches, here we propose a new benchmark problem for the HTS modelling community [20]: the HTS dynamo. This benchmark comprises a specific simplified geometry of an HTS dynamo, with well-defined inputs (i.e., assumptions) and an expected set of outputs (i.e., the solution). This allows any modelling technique to be validated against the expected solution, and its performance critically compared with other state-of-the-art methods for modelling superconductors.

In this work, this benchmark problem is implemented using several different methods:

- Coupled **H**-**A** formulation (H-A) [21];
- **H**-formulation + shell current (H+SC) [14, 15, 19];
- Segregated **H**-formulation (SEG-H) [22];
- Minimum Electromagnetic Entropy Production (MEMEP) [23, 24];
- Coupled **T**-**A** formulation (T-A) [25, 26];
- Integral equation (IE) [27];
- Volume integral equation-based equivalent circuit (VIE) [28].

Section 2 details the benchmark problem, including the geometry, parameters and the relevant assumptions. Section 3 describes each of the modelling frameworks, including how the open-circuit voltage is defined. Section 4 presents the open-circuit equivalent instantaneous voltage waveforms and the cumulative time-averaged equivalent voltages, as well as the current density and electric field distributions within the HTS wire during the magnet transit. Finally, in section 5, a critical analysis and comparison of each of the modelling frameworks is presented, based on the following key metrics: number of mesh elements in the HTS wire, total number of mesh elements in the model, number of degrees of freedom (DOFs), tolerance settings and the approximate time taken per cycle for each model.

2. The HTS dynamo benchmark

The geometry of the HTS dynamo benchmark problem is shown in Figure 1, assuming for simplicity the 2D (infinitely long) case. The permanent magnet (PM) rotates anticlockwise past the stationary HTS wire in the open-circuit configuration [14]. The PM has a width $a$ and height $b$ and a remanent flux density $B_r$. The initial position of the PM is such that the centre of its outer face is located at (0, –$R_{rotor}$), i.e., $\vartheta_M(t = 0) = -\pi/2$. The HTS wire has a width $e$ and thickness $f$ and is positioned such that its inner face is located (0, $R_{rotor}$ + airgap).

Table I lists the assumed parameters for the model, which are based on the model presented in [14] and correspond to the experimental setup in [29]. For simplicity, only the superconducting layer of the HTS wire is modelled and $J_c$ is assumed to be constant (where $J_c = I_c/(e \cdot f)$). It was shown in [14] that this assumption does not impact the essential dynamics to deliver a DC voltage, which is simply that the wire must exhibit a non-linear resistivity. Isothermal conditions are assumed (i.e., a constant temperature, $T$) and hence no thermal model needs to be included. The frequency of the PM rotation is 4.25 Hz, which was analysed in [14] using the H+SC method and compared with the experimental data taken at the same frequency in [29].

Regardless of the modelling framework used, the open-circuit equivalent instantaneous voltage and the cumulative time-averaged equivalent voltage waveforms shown later in Figures 3 and 4, respectively, should be obtained by implementing the benchmark.

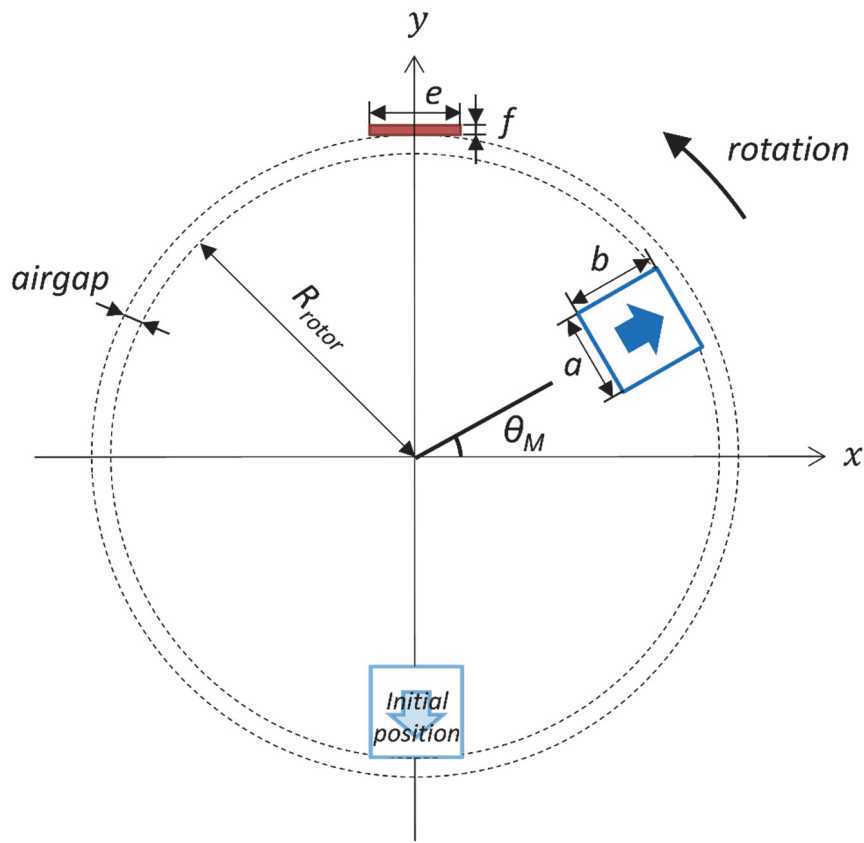

**Figure 1.** Geometry of the HTS dynamo benchmark problem. A permanent magnet (PM) rotates anticlockwise past an HTS wire.

**Table I.** HTS dynamo benchmark parameters.

| | | |
|---|---|---|
| **Permanent magnet (PM)** | *Width, a* | 6 mm |
| | *Height, b* | 12 mm |
| | *Active length (depth), L* | 12.7 mm |
| | *Remanent flux density, $B_r$* | 1.25 T |
| **HTS stator wire** | *Width, e* | 12 mm |
| | *Thickness, f* | 1 µm |
| | *Critical current, $I_c$ [self-field, 77 K]* | 283 A |
| | *n value* | 20 |
| *Rotor external radius, $R_{rotor}$* | | 35 mm |
| *Distance between PM face & HTS surface, airgap* | | 3.7 mm |
| *Frequency of rotation* | | 4.25 Hz |
| *Number of cycles* | | 10 |

3. Modelling frameworks

3.1    General definitions

The nonlinear resistivity, $\rho(J)$, of the superconductor is simulated using an *E-J* power law [30-32]:

$$\mathbf{E} = \frac{E_0}{J_c} \left| \frac{J}{J_c} \right|^{n-1} \mathbf{J} \tag{1}$$

where, in Cartesian coordinates and infinitely long (2D) problems in the *z* direction, $\mathbf{J} = [0\ 0\ J_z]$ and $\mathbf{E} = [0\ 0\ E_z]$ are the current density and electric field, respectively, which are assumed to be parallel to each other such that $\mathbf{E} = \rho\mathbf{J}$. $E_0 = 1$ µV/cm is the characteristic electric field and *n* defines the steepness of the transition between the superconducting state and the normal state. For *n* > 20, equation (1) becomes a reasonable approximation of the Critical State Model (CSM), for which *n* approaches infinity [33, 34], although accurate agreement with the CSM may require *n* values in the range of 100-1000 [24].

In general, the instantaneous measured voltage, *V*(*t*), is the path integral of the gradient of the electrostatic potential, $\nabla\varphi$, over the superconductor and measurement wires [14, 35]. When the excitations are periodic with period *T* (external magnetic field and transport currents) and for infinitely long geometries, the DC component of the voltage

$$V_{DC} = \frac{1}{T}\int_t^{t+T} V(t')\, dt' \tag{2}$$

corresponds to [14, 15, 18]

$$V_{DC} = -\frac{L}{T} \int_{t}^{t+T} E_{ave}(t') \, dt' \tag{3}$$

where $L$ is the active length of the dynamo, i.e., the active length (depth) of the PM, and $E_{ave}$ is the electric field, $E_z$, averaged over the cross-section of the superconductor, $S$:

$$E_{ave}(t) = \frac{1}{S} \iint_S E_z(x,y,t) \, dS \tag{4}$$

Then, the equivalent instantaneous voltage, $V_{eq}(t)$, is defined as

$$V_{eq}(t) = -L E_{ave}(t) \tag{5}$$

and the cumulative time-averaged equivalent voltage, $V_{cumul}(t)$, is given by

$$V_{cumul}(t) = \frac{1}{t} \int_0^t V_{eq}(t) \, dt \tag{6}$$

which corresponds to $V_{DC}$ for large enough $t$. Under open-circuit conditions, no net transport current flows, such that, at all times

$$I(t) = \iint_S J_z(x,y,t) \, dS = 0 \tag{7}$$

which is implemented as a constraint in each of the models.

### 3.2. **H**-formulation models

For the 2D **H**-formulation [36-42], the independent variables are the components of the magnetic field strength, **H** = [$H_x$ $H_y$ 0], and the governing equations are derived from Maxwell's equations – namely, Ampere's (8) and Faraday's (9) laws:

$$\nabla \times \mathbf{H} = \mathbf{J} \tag{8}$$

$$\nabla \times \mathbf{E} = -\frac{\partial \mathbf{B}}{\partial t} \tag{9}$$

The permeability $\mu = \mu_0$, and equations (8) and (9) are combined with the *E-J* power law, equation (1).

#### 3.2.1. Coupled **H**-**A** formulation (H-A)

The coupled **H**-**A** formulation, proposed for modelling superconducting rotating machines in [21], models the entire rotating model, with the **H**-formulation solved in a small region local to the HTS wire and the magnetic vector potential **A** solved elsewhere (thus, much of the model follows the usual construction for conventional rotating machines). In such a mixed-formulation model, careful attention must be paid to coupling variables across common boundaries between the **H** and **A** subdomains to maintain continuity: this is achieved by coupling, in weak form, the electric field from the **A**-formulation to the **H**-formulation and coupling the tangential components of the magnetic field from the **H**-formulation to the **A**-formulation, equivalent to a Neumann boundary condition [21].

In the implementation of this model here, a simplification is made by limiting the region that directly solves the *vector* fields associated with Maxwell's equations to a small region surrounding the conductive (current-carrying) subdomain, i.e., the **H**-formulation subdomain including the HTS wire. This allows most of the model to use the magnetic *scalar* potential, $V_m$, to calculate the PM field, for which the following magnetic flux conservation equation holds:

$$-\nabla \cdot \left( \mu \nabla V_m - B_r \right) = 0 \tag{10}$$

where $B_r$ is the remanent flux density (1.25 T for the PM assumed here – see Table I – and zero elsewhere). In the magnetic vector potential formulation, the magnetic flux density is defined as

$$\mathbf{B} = \nabla \times \mathbf{A} \tag{11}$$

and the electric field as

$$\mathbf{E} = \frac{-\partial \mathbf{A}}{\partial t} \tag{12}$$

automatically fulfilling Faraday's law (equation (9)) and the magnetic flux conservation law

$$\nabla \cdot \mathbf{B} = 0 \tag{13}$$

and then Ampere's law (equation (8)) is solved.

This model is implemented in COMSOL Multiphysics® using the Rotating Machinery, Magnetic (RMM) interface in the AC/DC module, which uses this mixed formulation of $V_m$ and **A**. An appropriate gauge is chosen such that the scalar electric potential (see equations (22) and (37)) vanishes and only **A** has to be considered (equation (12)). The **H**-formulation is implemented in the Magnetic Field Formulation (MFH) interface, also in the AC/DC module. In addition to the **H-A** coupling described above, the in-built Mixed Formulation Boundary node in the RMM interface imposes continuity between $V_m$ and **A** on either side of the magnetic scalar/vector potential boundary, such that

$$\mathbf{n}_1 \times \mathbf{H}_\mathbf{A} = \mathbf{n}_1 \times \mathbf{H}_{V_m} = \mathbf{n}_1 \times \left( -\nabla V_m \right) \tag{14}$$

and

$$\mathbf{n}_2 \cdot \mathbf{B}_{V_m} = \mathbf{n}_2 \cdot \mathbf{B}_A = \mathbf{n}_2 \cdot \nabla \times \mathbf{A} \tag{15}$$

where the surface normals, $\mathbf{n}_1$ and $\mathbf{n}_2$, are antiparallel ($\mathbf{n}_1 = -\mathbf{n}_2$). Equations (14) and (15) are implemented as weak contributions and can be interpreted as a surface current density, $\mathbf{J}_s = -\mathbf{n} \times \mathbf{H}$, and magnetic surface charge density, $\sigma_m = \mathbf{n} \cdot \mathbf{B}$, respectively.

### 3.2.2. H-formulation + shell current (H+SC)

Modelling the rotation of the PM in Fig. 1 is not trivial in the **H**-formulation. The approach developed in [14], and used again in [15, 19], is to represent the PM as a time-dependent sheet current $\mathbf{K}_{sheet}$ along the boundary of the rotor domain $\partial\Omega_R$. This reproduces the rotating field of the PM in the domain of interest (i.e., outside the rotor), while not introducing a rotating mesh or inter-model couplings to capture the rotation. The major advantage of this method is that it does not require the self-field correction described in Section 3.2.3 (segregated **H**-formulation), and ensures the whole model is solved natively as a finite element problem. However, the major disadvantage of this is the large number of mesh elements committed to simulating the boundary of the rotor domain which leads to a higher computational cost.

To implement the shell model, the PM is simulated in the sub-domain of the rotor $\Omega_R$ in a static model. Setting the boundary condition of the model to be magnetically insulating along the boundary of the rotor $\partial\Omega_R$:

$$\mathbf{n} \times \mathbf{E}\big|_{\partial\Omega_R} = 0 \tag{16}$$

This gives a solution where the flux of the magnet is completely enclosed in the rotor domain and **H** = 0 elsewhere. By the principle of superposition, it must be the case that the field of the PM is contained inside the boundary $\partial\Omega_R$ by a shell current, $\mathbf{K}_{shell}$, on this boundary, that produces the opposite PM field outside the boundary, i.e., $\mathbf{H}_m + \mathbf{H}_{shell} = 0$. Hence, the effect of the shell current is to produce the image of the PM's magnetic field outside the boundary of the initial model. The full problem can now be solved by omitting the original magnet and instead using the negative of the shell current distribution:

$$\nabla \times \mathbf{H}\big|_{\partial\Omega_R} = -\mathbf{K}_{shell}(\theta - \theta_M(t)) \tag{17}$$

where $\vartheta$ is the angular coordinate around the rotational axis of the rotor. This produces the magnetic field of the PM, in the domain of interest, and rotates it by the angle $\vartheta_M(t)$ with time.

### 3.2.3. Segregated H-formulation (SEG-H)

The segregated model is comprised of a *magnetostatic* PM model and a *time-dependent* **H**-formulation HTS wire model. The former is coupled unidirectionally to the latter using boundary conditions [22] and a translation (rotation) operator for the PM's static magnetic field (see Figure 2). This avoids, like the preceding shell current model, the need to model the rotating PM (e.g., using a moving mesh), but also significantly reduces the number of mesh elements in the HTS model.

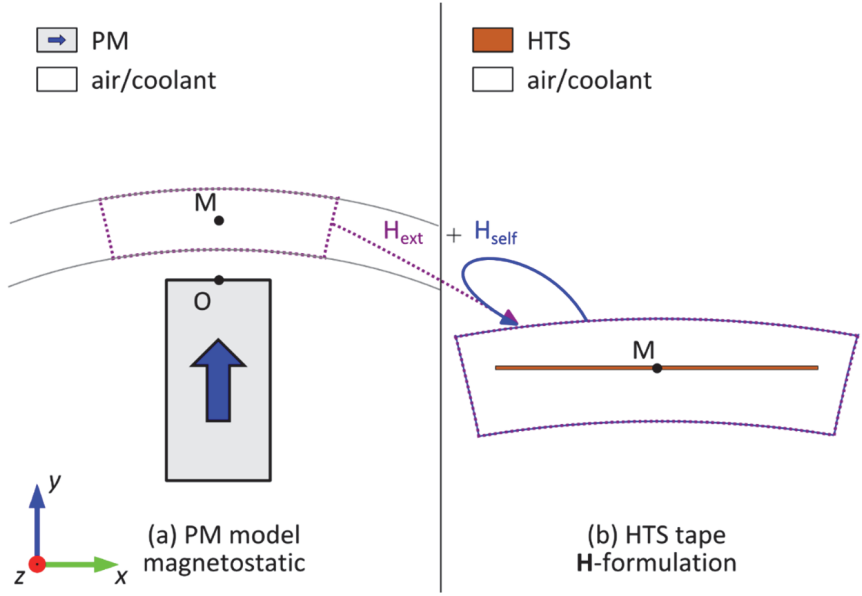

**Figure 2.** Segregated finite-element model: (a) *magnetostatic* PM model, (b) *time-dependent* **H**-formulation HTS wire model.

On the outer boundary of the **H**-formulation model, the sum of the applied field, **H**$_{ext}$, and the self-field, **H**$_{self}$, is applied as a Dirichlet boundary condition. To mimic the rotation, **H**$_{ext}$ is obtained by rotating the field of a static permanent magnet, **H**$_{PM}$:

$$\begin{bmatrix} H_{ext,x}(x,y,t) \\ H_{ext,y}(x,y,t) \end{bmatrix} = \begin{bmatrix} \cos\theta_M(t) & \sin\theta_M(t) \\ -\sin\theta_M(t) & \cos\theta_M(t) \end{bmatrix} \begin{bmatrix} H_{PM,x}(x_{rot},y_{rot}) \\ H_{PM,y}(x_{rot},y_{rot}) \end{bmatrix} \quad (18)$$

$$\begin{bmatrix} x_{rot} \\ y_{rot} \end{bmatrix} = \begin{bmatrix} \cos\theta_M(t) & -\sin\theta_M(t) \\ \sin\theta_M(t) & \cos\theta_M(t) \end{bmatrix} \begin{bmatrix} x \\ y \end{bmatrix} \quad (19)$$

where $\vartheta_M(t)$ is the rotor angle and ($x_{rot}$, $y_{rot}$) are the coordinates in the rotated coordinate system.

The self-field, **H**$_{self}$, created by the supercurrent flowing in the HTS wire, is calculated at each time step by numerical integration of the 2D Biot-Savart law over the HTS wire subdomain:

$$H_{self,x}(x,y,t) = \frac{1}{2\pi} \iint_S \frac{-J_z(x',y',t) \cdot (y-y')}{(x-x')^2 + (y-y')^2} dx'dy' \quad (20)$$

$$H_{self,y}(x,y,t) = \frac{1}{2\pi} \iint_S \frac{J_z(x',y',t) \cdot (x-x')}{(x-x')^2 + (y-y')^2} dx'dy' \quad (21)$$

### 3.3 Minimum Electromagnetic Entropy Production (MEMEP)

MEMEP is a variational method that has been shown to be ideally suited modelling materials with highly nonlinear **E(J)** relationships, such as superconductors [23, 24]. MEMEP solves the current density **J** by minimizing a functional that contains all the variables of the problem, including the magnetic vector potential **A**, current density **J** and scalar potential $\varphi$. It has been proven that this functional always presents a minimum, it is unique, and it is the solution of Maxwell's equations in differential form [24].

This method is fast because the current density only exists inside the superconducting region, and thus the mesh is only required inside this region. The general equation for the current density and the scalar potential are

$$\mathbf{E} = -\frac{\partial \mathbf{A}}{\partial t} - \nabla \varphi \qquad (22)$$

$$\nabla \cdot \mathbf{J} = 0 \qquad (23)$$

In Coulomb's gauge ($\nabla \cdot \mathbf{A} = 0$), **A** can be separated into the contributions from the current density, $\mathbf{A}_J$, and the external sources, $\mathbf{A}_a$, [23]. For infinitely long problems in the *z* direction (or 2D), **J** = [0 0 $J_z$], **E** = [0 0 $E_z$], and **A** = [0 0 $A_z$]. The $A_J$ contribution in Coulomb's gauge follows [33]

$$A_J(r) = -\frac{\mu_0}{2\pi} \int_S dS' J(r') \ln|r - r'| \qquad (24)$$

Equation (23) is always satisfied for 2D problems, and thus only equation (22) needs to be solved. To solve this equation, the following functional should be minimised [23, 24]:

$$L = \int_S ds \left[ \frac{1}{2} \frac{\Delta A_J}{\Delta t} \cdot \Delta J + \frac{\Delta A_a}{\Delta t} \cdot \Delta J + U(J_0 + \Delta J) \right] \qquad (25)$$

where *U* is the dissipation factor defined as [24]

$$U(J) = \int_0^J E(J') \cdot dJ' \qquad (26)$$

This dissipation factor can include any *E-J* relationship for superconductors, including the multi-valued relation of the CSM [24, 43, 44]. In this problem, the non-uniform applied magnetic field caused by the rotating PM appears in the functional in the form of $A_a$. Then, the program only requires calculation of the vector potential once for each time step within the first cycle. The impact on the total computing time is negligible because the minimisation takes most of the computing time.

The vector potential generated by the PM with uniform magnetisation **M** can be calculated by the magnetization sheet current density **K** = **M** × $\mathbf{e}_n$, where **M** is the PM magnetisation and $\mathbf{e}_n$ is the unit vector normal to the surface. For uniform magnetisation, the vector potential $\mathbf{A}_M$ generated by the PM is:

$$A_M(r) = -\frac{\mu_0}{2\pi} M \int_{\partial S} dl' \mathbf{e}_m \times \mathbf{e}_n(r') \ln|r - r'| \qquad (27)$$

where $\mathbf{e}_m$ is the unit vector in the magnetisation direction, *∂S* represents the edges of the PM cross-section, and *dl'* is the length differential on the edge. In this work, $\mathbf{A}_M$ was evaluated numerically. Note that the cross product in the equation above always follows the *z* direction, since both $\mathbf{e}_m$ and $\mathbf{e}_n$ are in the *xy* plane.

MEMEP can also take a $J_c(B, \vartheta)$ dependence into account by solving **J**, then calculating **B**, and iterating until the difference is below a certain tolerance [23]. In this case, **B** from the PM should also be calculated, as carried out in [18].

### 3.4 Coupled **T-A** formulation (T-A)

The **T-A** formulation was proposed in [25, 45] to tackle the problem of simulating superconductors characterised by a very high aspect (width:thickness) ratio, such as HTS wires. The main idea is to use the magnetic vector potential **A** for calculating the magnetic field in the whole domain and the current vector potential **T** for calculating the current density

$$\mathbf{J} = \nabla \times \mathbf{T} \tag{28}$$

in the superconductor. The obtained current density is re-injected as an external current density in the **A**-formulation part.

The superconductor can be simulated as a 1D object in this 2D problem, the 1D line representing the superconducting wire. Further resulting simplifications are that the current vector potential has only one component and that the transport current flowing in the superconductor can be imposed by means of simple boundary conditions for **T** at the wire's edges. In particular, the current is determined by the difference between the values of **T** at the wire's edges. In the benchmark considered here, there is no transport current and the simple condition **T** = *any constant* is applied at each of the wire's edges.

The **T-A** formulation has been recently extended to simulate superconductors of finite thickness [46]. In this case, **T** has two components and the setting of the boundary conditions is a little less intuitive – see [46] for details. To distinguish between these two **T-A** formulations, we refer to these hereafter as **T-A** (1D) and **T-A** (2D), respectively. In addition, these two formulations are implemented separately using only the magnetic vector potential **A** (implemented in COMSOL using the Magnetic Field (MF) interface), referred to as VP (vector potential), and using the mixed scalar-vector potential detailed in Section 3.2.1, implemented using the Rotating Machinery, Magnetic (RMM) interface and referred to as SP (scalar potential).

### 3.5 Integral equation (IE)

The current distribution along a segment representative of an infinitely thin (1D) superconducting layer (as per Section 3.4) can be given by an integral equation that can be easily solved by the finite-element method. This approach was proposed in [27], and later extended to consider interacting HTS wires in [47, 48].

For the problem analysed here, the integral equation is written in the Partial Differential Equation (PDE) module of COMSOL in 1D and takes the following form

$$\rho J_s = \mu f (Q + K) + C \tag{29}$$

with

$$K(x,t) = \int_{-a}^{x} \partial_t H_n(\xi, t)\, d\xi \tag{30}$$

and

$$Q(x,t) = \frac{1}{2} \int_{-a}^{a} \partial_t J_s(\xi, t) \ln|\xi - x|\, d\xi \tag{31}$$

where $\rho$ is the power-law resistivity (equation (1)), $J_s$ is the sheet current density (current per unit width, A/m) in the $z$ direction, $f$ is the thickness of the superconductor (see Figure 1), $a$ is the half-width (i.e., *e* = 2*a*) and $H_n$ is the normal component of the external field impinging on the superconductor. The constant C is set at each time step to satisfy the constraint on the desired transport current, which in this benchmark is always zero (see equation (7)):

$$\int_{-a}^{a} J_s(\xi,t) d\xi = 0 \tag{32}$$

To obtain the external field, the 1D PDE module is coupled to the 2D Magnetic Field (MF) interface in the AC/DC module, which calculates only the magnetic field generated by the rotating PM. The mesh rotation is considered by using the arbitrary Lagrangian-Eulerian (ALE) method.

The MF module does not contain the "reaction" term of the field created by the currents flowing in the superconductor. As a consequence, in order to visualize the total magnetic field in the whole simulated domain, one needs to use a second MF module where the magnetic field generated by the sheet current is added to the one generated by the permanent magnet. This has, of course, an additional computational cost because of the additional DOFs. For the benchmark proposed here, where the focus is the quantities in the superconductor, this second MF module was not used.

The main advantage of this method relies in the fact that the current density is the state variable of the equation and, as a consequence, it is not obtained by the spatial derivative as required by Ampere's law (equation (8)).

Note that in equations (29)-(31), the time derivative of $J_s$ appears under the integral sign, but using standard procedure (Carleman's equation), it is possible reformulate these in a way that extracts the time derivative:

$$\partial_t J_s(x,t) = \left(\frac{2}{\pi \mu f}\right) \int_{-a}^{a} \frac{\partial_\xi F(\xi,t)}{\xi - x} \sqrt{\frac{a^2 - \xi^2}{a^2 - x^2}} d\xi + C(t) \tag{33}$$

where $\quad F(x,t) = \rho J_s(x,t) - \mu f K(x,t) \tag{34}$

Therefore,

$$\partial_t J_s(x,t) = \frac{2}{\pi} \int_{-a}^{a} \frac{\partial_\xi F(\xi,t)}{\xi - x} \sqrt{\frac{a^2 - \xi^2}{a^2 - x^2}} d\xi + C(t) \tag{35}$$

where $\quad F(\xi,t) = (\frac{\rho}{\mu f}) J_s(\xi,t) - K(\xi,t) = \frac{E(\xi,t)}{\mu} - K(\xi,t) \tag{36}$

This second form may be more appropriate for standard numerical routines for solving integral equations (Nystroem method) that do not require finite elements on the strip segment.

3.6    Volume integral equation-based equivalent circuit (VIE)

The volume integral equation-based equivalent circuit is obtained by separating the total electromotive force at any point of the superconductor into two contributions: one contribution due to the time-varying field produced by the current induced in the superconductor and a second contribution due to the movement of the PM. According to this, and by expressing the magnetic flux density via the magnetic vector potential as in equation (11), Faraday's law at any point in the superconductor gives [49-51]

$$\mathbf{E} = -\frac{\partial \mathbf{A}^{int}}{\partial t} - \mathbf{v} \times \mathbf{B}^{PM} - \nabla \varphi \qquad (37)$$

where $\mathbf{A}^{int}$ is the vector potential of the current in the superconductor, $\mathbf{B}^{PM}$ is the PM field and $\mathbf{v}$ is the velocity of $\mathbf{B}^{PM}$ at the considered point, expressed in the fixed reference frame of Fig. 1. A numerical solution of the problem is obtained by subdividing the superconductor domain into a finite number of 2D elements and by enforcing equation (37) to be satisfied, in the weak form, over each element of the discretization [28, 52]. The state variables of the problem are the current densities of each element. This is obtained by relating $\mathbf{E}$ and $\mathbf{A}^{int}$ in equation (37) to the current density via the *E-J* power law (equation (1)) and equation (24), respectively. The thin shell model (see Sections 3.2.2 and 3.3) is used for calculating the field of the PM at each position. Each of the discretised equations obtained via the weighted residual approach corresponds to the voltage balance of a circuit branch involving a non-linear resistor arising from the electric field in the superconductor, a coupled inductor representing the magnetic interaction of the induced current, and a voltage generator corresponding to the Lorentz-like electromotive term. Hence, this circuit picture gives rise to the name of the method.

4. Results

Figure 3 shows a comparison of the open-circuit equivalent instantaneous voltage waveforms calculated by each of the models for the 2nd transit of the PM past the HTS wire, ignoring any initial transient effects that may be present in the 1st cycle. Indeed, qualitatively, the distinct four peaks and noticeable left-to-right asymmetry observed in experiments (see [14], for example) are reproduced and there is excellent quantitative agreement between the models for the magnitude of these peaks. The particular characteristics of this voltage waveform give rise to the DC output voltage of the HTS dynamo, which can be further evidenced by examining the cumulative time-averaged equivalent voltage, $V_{cumul}$, given by equation (6). $V_{cumul}$ calculated by each of the models over the 10 cycles of PM rotation is shown in Figure 4. In all cases this converges to a non-zero asymptotic value, clearly showing the DC output.

Again there is excellent qualitative and quantitative agreement between each of the models: the average value of $V_{cumul}$ is –9.41 µV with a standard deviation of 0.34 µV. It should be noted that, as described earlier, there is some discrepancy between these results and those observed in experiments because of the use of the constant $J_c$ approximation. The use of the angular field-dependence of $J_c(B, \vartheta)$ to consider the suppression of $J_c$ with magnetic field, as considered elsewhere in [14, 15, 18, 19], is needed for good agreement with experiment and can be done easily by modifying the *E-J* power law, equation (1), such that $J_c = J_c(B, \vartheta)$.

Figure 5 shows the current density normalised to $J_{c0}$, $J/J_{c0}$, and electric field, $E$, distributions within the wire for three key PM positions as the magnet travels past it:

1) as the magnet approaches the centre of the wire from the right-hand side ($t$ = 347 ms in the 2nd cycle);

2) when the magnet is aligned with the centre of the wire ($t$ = 353 ms); and

3) as the magnet moves away from the wire on its left-hand side ($t$ = 359 ms).

The dynamics of the current flowing within the wire and the related local electric fields ultimately give rise to the voltage waveforms shown in Figure 3, with 1) close to the first negative peak, 2) close to the trough in between the 2nd and 3rd (positive) peaks and 3) close to the fourth negative peak. At 1), an overcritical ($J > J_c$) eddy current flows in the right-hand side of the wire, which then returns (i.e., flows in the opposite direction) on the other side of the wire at a lower magnitude, giving rise to the asymmetric electric field distribution as described by the *E-J* power law. At 2), these forward and reverse currents are almost equal and opposite, such that V ≈ 0, and at 3), the reverse situation to 1) occurs, such that the *J* and *E* profiles are essentially mirrored. Similar profiles were also obtained in [14] for the $J_c(B, \vartheta)$ case, except that the suppression of $J_c$ with magnetic field results in higher local electric fields in the region close to the PM.

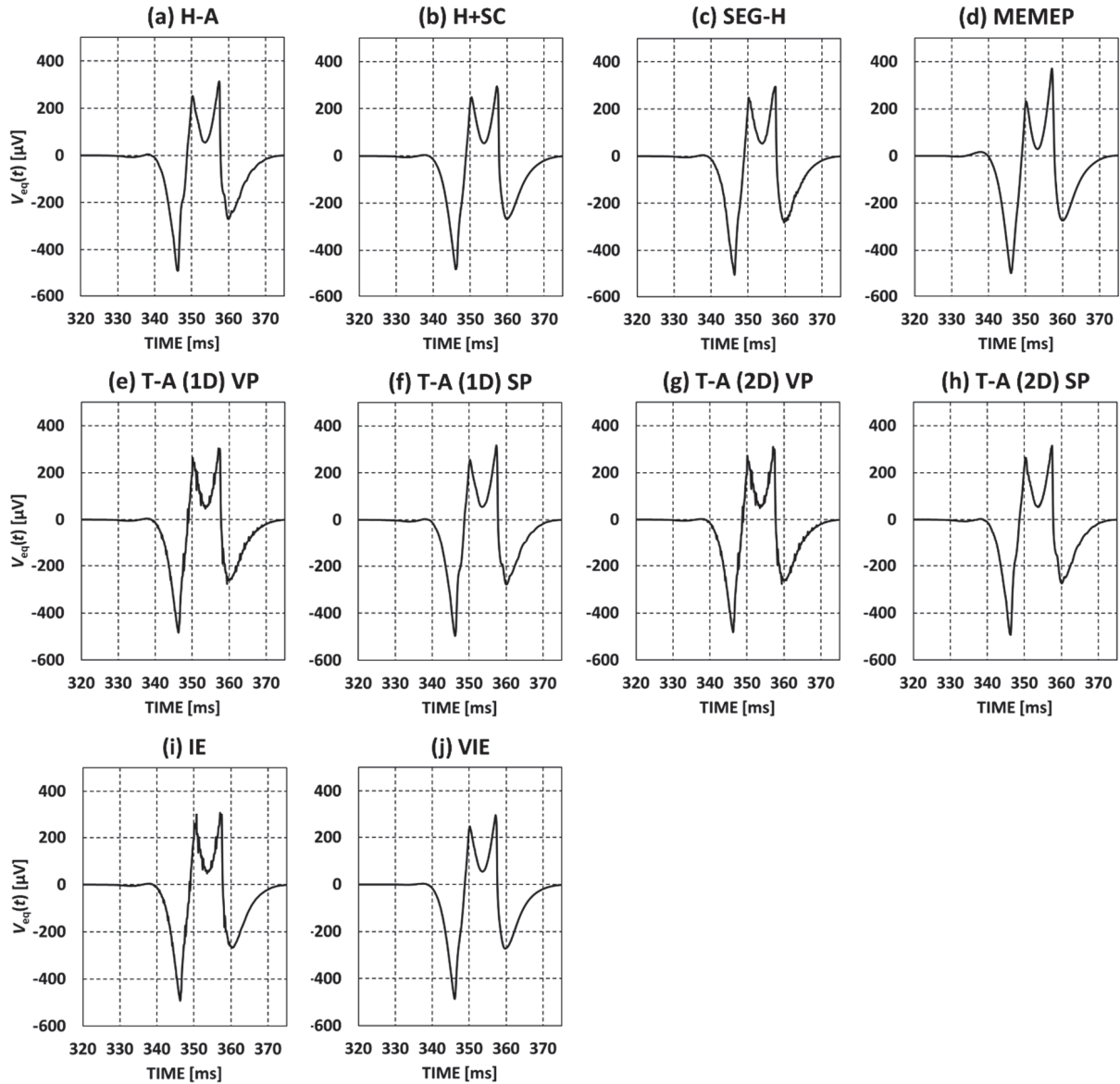

**Figure 3.** Open-circuit equivalent instantaneous voltage, $V_{eq}(t)$, waveforms calculated by each of the models for the 2nd transit of the PM past the HTS wire, ignoring any initial transient effects that may be present in the 1st cycle. Qualitatively, the distinct four peaks and noticeable left-to-right asymmetry observed in experiments are reproduced and there is excellent quantitative agreement between the models for the magnitude of these peaks.

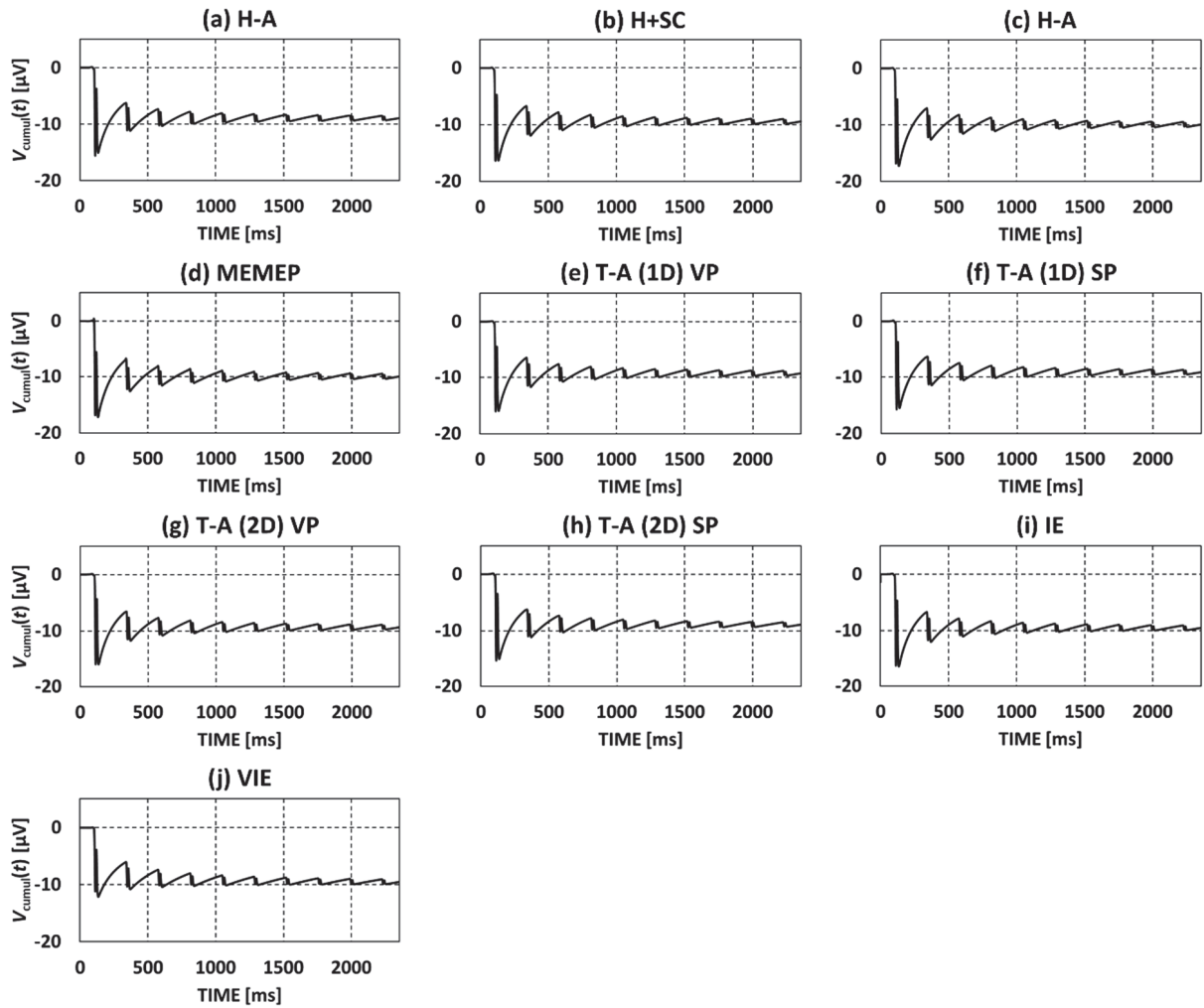

**Figure 4.** Cumulative time-averaged equivalent voltage, $V_{cumul}(t)$, calculated for each of the models over the 10 cycles of PM rotation, clearly showing a DC output, with excellent qualitative and quantitative agreement between each of the models. The average value of $V_{cumul}$ after 10 cycles of PM rotation is –9.41 µV with a standard deviation of 0.34 µV.

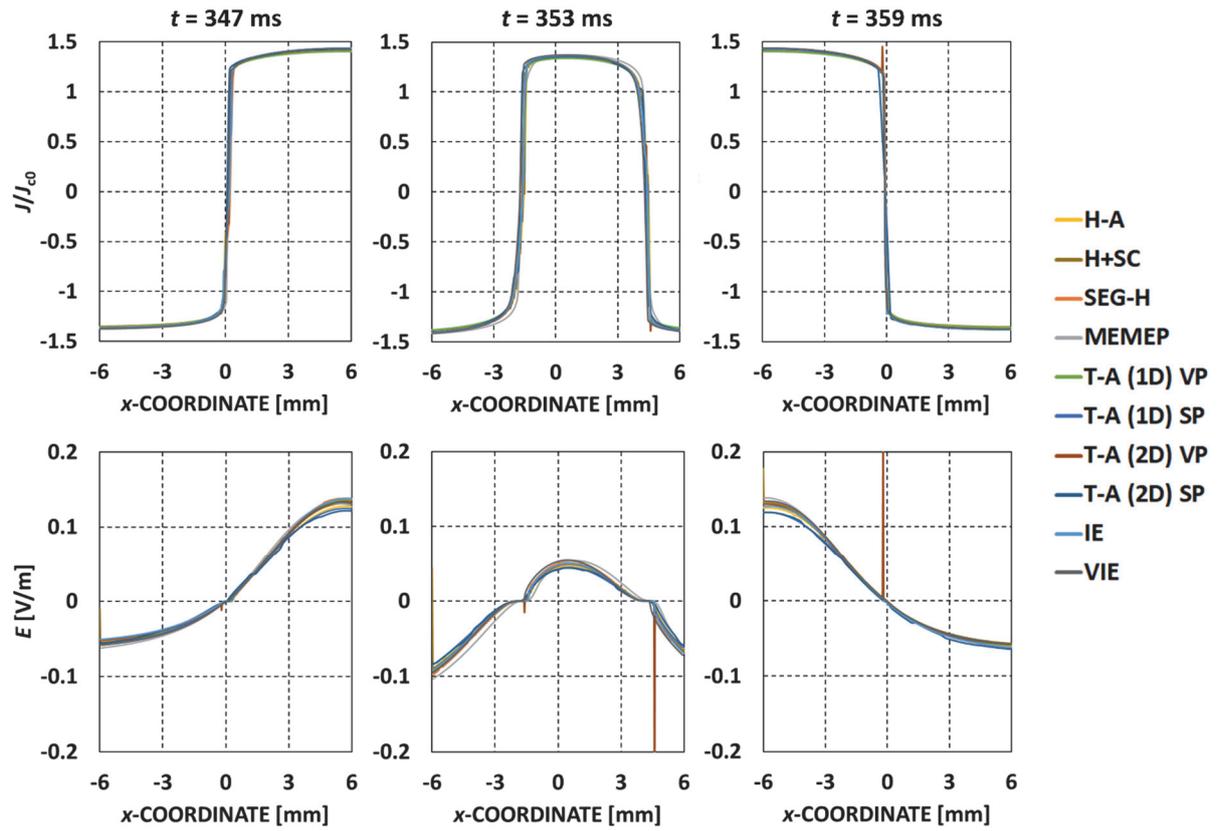

**Figure 5.** Distributions within the wire for: current density normalised to $J_{c0}$, $J/J_{c0}$, and electric field, $E$. The calculated distributions are shown for three key PM positions: 1) as the magnet approaches the centre of the wire from the right-hand side ($t$ = 347 ms in the 2nd cycle); 2) when the magnet is aligned with the centre of the wire ($t$ = 353 ms); and 3) as the magnet moves away from the wire on its left-hand side ($t$ = 359 ms). The dynamics of the current flowing within the wire and the related local electric fields ultimately give rise to the voltage waveforms shown in Figure 3.

5. Discussion

As shown in Figures 3-5, all of the models produced the expected benchmark solution with excellent qualitative and quantitative agreement. In this section, a critical analysis and comparison of each of the modelling frameworks is presented. Table II lists the key metrics assessed for each benchmark model: the number of mesh elements in the HTS wire; the total number of mesh elements in the model; the number of DOFs; the relative and absolute tolerance settings (for FEM-based models), tolerance for the mutual interaction matrix (MEMEP; programmed in C++) or *ode23b* solver relative tolerance (VIE; programmed in MATLAB); and the approximate time taken per cycle for each model. In the interest of a fair comparison, all of the FEM-based models were run on the same computer under the same conditions (e.g., COMSOL Multiphysics™ version 5.5) and, where possible, the number of mesh elements in the HTS wire set to 120 x 1 along the width and thickness, respectively. The key findings and comparisons are detailed below:

- The clear winner in terms of computational speed is the MEMEP method, with the entire 10 cycles taking a little over two minutes to solve. This can be explained by the limited number of DOFs because only the HTS wire needs to be meshed. It should also be noted that the MEMEP model was run on a slightly inferior processor, so slightly overestimates the computational time per cycle. The next best performers are the SEG-H and VIE methods, which are also modelling frameworks that emphasise a reduced number of mesh elements.
- It should be noted that, in the FEM-based models, the mesh was optimised as best possible as a compromise between accuracy and computational speed. Thus, there is scope in many of the FEM-based models to improve their accuracy somewhat by using a finer mesh, but there will be an associated increase in computational time. Several models (H-A, T-A (2D), SEG-H and H+SC) took advantage of the artificial expansion technique presented in [53, 54] to increase the HTS layer thickness from 1 µm to 100 µm, improving the computational speed without compromising accuracy.
- The rotating machine-like models that made use of the mixed scalar-vector potential (H-A, T-A (1D) SP and T-A (2D) SP) also performed well in terms of computational time, and the integral equation-based model (IE), even with the use of the magnetic vector potential with second-order (quadratic) elements and the associated significant increase in DOFs, also performed comparably.
- The T-A (2D) formulations – both SP and VP – are found to be reasonably unstable, and even a finer mesh (60 x 4) and tighter tolerances settings could not improve this performance dramatically. The spikes seen in the *J* and *E* plots in Figure 5 are from the T-A (2D) VP model; the *E* spikes are much more pronounced due to *E* being proportional to $J^n$ (see equation (1)). Spurious oscillations of a similar kind were also presented in [45, 55] for the T-A (1D) model used for AC loss calculations. It was recommended in both [45, 55] that second-order (quadratic) elements be used for **A** to mitigate this, which does have its associated computational cost. However, it should also be noted that for the benchmark here and in [45, 55], this did not impact the calculations of interest significantly (AC loss and voltage, respectively), although the voltage waveforms in Figure 3 are clearly noisier in comparison to other models.
- The use of the scalar potential (H-A, T-A (1D) SP and T-A (2D) SP) – with first-order (linear) elements – improved stability and spurious oscillations/noise, as well as computational speed, in comparison to the VP models. Using this mixed scalar-vector formulation is a potentially useful alternative for modelling such dynamos, as well as superconducting rotating machines in general. For 3D models in particular, the scalar potential formulation introduces fewer DOFs and can ensure a more accurate coupling of the magnetic field. Indeed, although all of these models were specifically created for the HTS dynamo benchmark, many of the findings are equally applicable to and useful for modelling superconducting rotating machines.
- In terms of ease of use, all models (except for the MEMEP and VIE methods, which are self-programmed using C++ and MATLAB, respectively) were implemented in COMSOL Multiphysics™. COMSOL is a popular commercial software package with a reasonably gentle learning curve and is currently used by dozens of research groups worldwide to model superconductivity-related problems [56, 57]. Many shared modelling examples, most associated with peer-reviewed publications, are available on the HTS Modelling Workgroup website [58]. COMSOL now has a dedicated superconductivity interface (the MFH interface; see Section 3.1.1) with dedicated technical support.
- However, there is a significant disadvantage when using commercial software in that users do not have complete control over its implementation: some of the programming cannot be accessed easily, if at all. It is of particular note that the SEG-H and H+SC models, which were built and optimised in COMSOL version 5.4, ran significantly slower when opened and run in version 5.5. The computational time per cycle for both versions is included in Table II. At the time of writing, COMSOL were unable to explain why these models had longer run times in version 5.5, despite backward compatibility. There is also an associated financial cost that can be a barrier to some researchers, which is where self-programmed techniques and those implemented in free software have a distinct advantage.

**Table II.** Key metrics assessed for each benchmark model.

| Model | Mesh (SC) | Mesh (total) | DOFs | Rel./abs. tolerance | Approx. time/cycle [min/cycle] | Software implementation |
|---|---|---|---|---|---|---|
| MEMEP | 120 (120 x 1) | 120 | 120 | 1e-4[1] | < 0.25[A] | C++ |
| SEG-H | 120 (120 x 1) | 2653 | 4071 | 1e-4 / 0.1 | 1.1[B] | COMSOL 5.4 |
| | | | | | 2.6[B] | COMSOL 5.5 |
| VIE | 120 (120 x 1) | 120 | 120 | 1e-3[2] / 1e-6[2] | 1.6[B] | MATLAB |
| H-A | 120 (120 x 1) | 4176 | 3661 | 1e-4 / 0.1 | 2.1[B] | COMSOL 5.5 |
| T-A (2D) SP | 240 (60 x 4) | 3800 | 2863 | 1e-5 / 1e-4 | 3.9[B] | COMSOL 5.5 |
| IE | 120 (120 x 1) | 5932 | 12451 | 5e-3 / 0.1 | 5.1[B] | COMSOL 5.5 |
| T-A (1D) SP | 120 (120 x 1) | 4876 | 2779 | 1e-5 / 1e-4 | 6.5[B] | COMSOL 5.5 |
| H+SC | 120 (120 x 1) | 11272 | 16988 | 1e-5 / 1e-3 | 7.9[B] | COMSOL 5.4 |
| | | | | | > 120 | COMSOL 5.5 |
| T-A (1D) VP | 120 (120 x 1) | 6064 | 12715 | 1e-4 / 0.1 | 21.6[B] | COMSOL 5.5 |
| T-A (2D) VP | 240 (60 x 4) | 5286 | 13696 | 1e-4 / 0.1 | 64.6[B] | COMSOL 5.5 |

PC specifications:
[A]Intel® Core™ i7-8700 CPU @ 3.20 GHz, 31.1 GB RAM (10% memory used for MEMEP model), Ubuntu 16.04 LTS, 64-bit
[B]Intel® Core™ i9-7900X CPU @ 3.30 GHz, 63.7 GB RAM, Microsoft Windows 10 Pro, 64-bit
Other notes:
[1]Tolerance for the mutual interaction matrix
[2]Default settings for MATLAB/*ode23b* solver

6. Conclusion

In this work, a new benchmark problem for the HTS modelling community – the HTS dynamo – was proposed, consisting of a permanent magnet (PM) rotating past a stationary HTS wire in the open-circuit configuration. The benchmark was then implemented using several different methods, including **H**-formulation-based methods, coupled **H**-**A** and **T**-**A** formulations, the Minimum Electromagnetic Entropy Production method, and integral equation and volume integral equation-based equivalent circuit methods.

Excellent qualitative and quantitative agreement was obtained between all models for the open-circuit equivalent instantaneous voltage and the cumulative time-averaged equivalent voltage, as well as the current density and electric field distributions within the HTS wire at key positions during the magnet transit. The average value for all the models of the DC output voltage of the HTS dynamo, determined by the cumulative time-averaged equivalent voltage over 10 cycles of PM rotation, was calculated to be $-9.41$ µV with a standard deviation of $0.34$ µV.

A critical analysis and comparison of each of the modelling frameworks was presented, based on the following key metrics: number of mesh elements in the HTS wire, total number of mesh elements in the model, number of DOFs, tolerance settings and the approximate time taken per cycle for each model. The clear winner in terms of computational speed is the MEMEP method, with the entire 10 cycles taking around two minutes to solve, due to the limited number of DOFs because only the HTS wire needs to be meshed. The next best performers were the SEG-H and VIE methods, which are also modelling frameworks that emphasise a reduced number of mesh elements. Several models took advantage of an artificial expansion technique to increase the HTS layer thickness from 1 µm to 100 µm, improving the computational speed without compromising accuracy.

A number of models use a rotating machine-like modelling framework – in particular, the coupled **H**-**A** and **T**-**A** formulations – and it is shown that the use of a mixed scalar-vector potential (implemented using COMSOL's Rotating Machinery, Magnetic interface) results in a significant improvement in both computational speed and stability, compared to models that use only the vector potential (implemented using COMSOL's Magnetic Field interface). In the latter case, it is recommended to use second-order (quadratic) elements for **A** – in particular, for the **T**-**A** formulation – to mitigate against spurious oscillations and improve stability, which has an associated computational cost. Using the mixed scalar-vector formulation provides a potentially useful alternative for modelling such dynamos, as well as superconducting rotating machines in general.

This benchmark and the results contained herein provide researchers with a suitable framework to validate, compare and optimise their own methods for modelling the HTS dynamo.

Acknowledgements

MA would like to acknowledge financial support from an Engineering and Physical Sciences Research Council (EPSRC) Early Career Fellowship EP/P020313/1. CB, RM and MA were supported in part by NZ Royal Society Marsden Grant No. MFP-VUW1806. FP would like to acknowledge the support of the Consejo Nacional de Ciencia y Tecnologia and Secretaria de Energia de Mexico, Reference No. 541016/439167. Additional data related to this publication are available at the University of Cambridge data repository (https://doi.org/10.17863/CAM.54005).

ORCID iDs

| | |
|---|---|
| Mark Ainslie | 0000-0003-0466-3680 |
| Francesco Grilli | 0000-0003-0108-7235 |
| Loïc Quéval | 0000-0003-3934-4372 |
| Enric Pardo | 0000-0002-6375-4227 |
| Fernando Perez-Mendez | 0000-0001-7980-9539 |
| Ratu Mataira | 0000-0003-0892-5430 |
| Antonio Morandi | 0000-0002-1845-4006 |
| Asef Ghabeli | 0000-0001-9907-4509 |
| Chris Bumby | 0000-0001-8555-2469 |
| Roberto Brambilla | |

References

[1] Hoffmann C, Pooke D and Caplin A D 2011 *IEEE Trans. Appl. Supercond.* 21 1628-31
[2] Volger J and Admiraal P 1962 *Phys. Lett.* 2 257-9
[3] Beelen H V *et al* 1965 *Physica* 31 413-43
[4] Bai Z *et al* 2010 *Cryogenics* 50 688-92
[5] Bumby C W *et al* 2016 *Supercond. Sci. Technol.* 29 024008
[6] Geng J *et al* 2016 *Appl. Phys. Lett.* 108 262601
[7] Geng J *et al* 2016 *J. Phys. D* 49 11LT01
[8] Bumby C W *et al* 2016 *Appl. Phys. Lett.* 108 122601
[9] Campbell A M 2017 *Supercond. Sci. Technol.* 30 125015
[10] Wang W and Coombs T 2018 *Phys. Rev. Appl.* 9 044022
[11] Giaever I 1966 *IEEE Spectrum* 3 117-22
[12] Kaplunenko V, Moskvin S and Schmidt V 1985 *Fiz. Nizk. Temp.* 11 846 [available at: https://fnte.ilt.kharkov.ua/fnt/pdf/11/11-8/f11-0846r.pdf]
[13] van de Klundert L and ten Kate H 1981 *Cryogenics* 21 195
[14] Mataira R C, Ainslie M D, Badcock R A and Bumby C W 2019 *Appl. Phys. Lett.* 114 162601
[15] Mataira R *et al* 2020 *Phys. Rev. Appl.* at press [https://journals.aps.org/prapplied/accepted/38073AfaE201f606317b4f8514f4729f71d101d20]
[16] Jiang Z *et al* 2014 *Appl. Phys. Lett.* 105 112601
[17] Vysotsky V S *et al* 1990 *Supercond. Sci. Technol.* 3 259-62
[18] Ghabeli A and Pardo E 2020 *Supercond. Sci. Technol.* 33 035008
[19] Mataira R, Ainslie M D, Badcock R and Bumby C W 2020 *IEEE Trans. Appl. Supercond.* 30 5204406
[20] HTS Modelling Workgroup: Benchmarks [http://www.htsmodelling.com/?page_id=2]
[21] Brambilla R *et al* 2018 *IEEE Trans. Appl. Supercond.* 28 5207511
[22] Quéval L *et al* 2018 *Supercond. Sci. Technol.* 31 084001
[23] Pardo E, Ŝouc J and Frolek L 2015 *Supercond. Sci. Technol.* 28 044003
[24] Pardo E and Kapolka M 2017 *J. Comput. Phys.* 344 339-63
[25] Zhang H *et al* 2017 *Supercond. Sci. Technol.* 30 024005
[26] Benkel T *et al* 2020 *IEEE Trans. Appl. Supercond.* 30 5205807
[27] Brambilla R *et al* 2008 *Supercond. Sci. Technol.* 31 105008
[28] Morandi A and Fabbri M 2015 *Supercond. Sci. Technol.* 28 024004
[29] Badcock R A *et al* 2017 *IEEE Trans. Appl. Supercond.* 27 5200905
[30] Plummer C J G and Evetts J E 1987 *IEEE Trans. Magn.* 23 1179-82
[31] Rhyner J 1993 *Physica C* 212 292-300
[32] Brandt E H 1997 *Phys. Rev. B* 55 14513-26

[33] Ainslie M D and Fujishiro H 2019 *IOP Expanding Physics* (Bristol: IOP Publishing) [DOI: 10.1088/978-0-7503-1332-2ch2] p 2-9
[34] Grilli F *et al* 2014 *IEEE Trans. Appl. Supercond.* 24 8200433
[35] Clem J R 1970 *Phys. Rev. B* 1 2140-55
[36] Kajikawa K *et al* 2003 *IEEE Trans. Appl. Supercond.* 13 3630-3
[37] Pecher R *et al* 2003 *Proc. 6th EUCAS* pp 1–11
[38] Hong Z, Campbell A M and Coombs T A 2006 *Supercond. Sci. Technol.* 19 1246-52
[39] Brambilla R, Grilli F and Martini L 2007 *Supercond. Sci. Technol.* 20 16-24
[40] Ainslie M D, Flack T J, Hong Z and Coombs T A 2011 *Int. J. Comput. Math. Electr. Electron. Eng.* 30 762-74
[41] Ainslie M D *et al* 2011 *IEEE Trans. Appl. Supercond.* 21 3265-8
[42] Ainslie M D, Flack T J and Campbell A M 2012 *Physica C* 472 50-6
[43] Bossavit A 1994 *IEEE Trans. Magn.* 30 3363-6
[44] Prigozhin L 1997 *IEEE Trans. Appl. Supercond.* 7 3866-73
[45] Liang F *et al* 2017 *J. Appl. Phys.* 122 043903
[46] Grilli F *et al* 2020 [https://arxiv.org/abs/2004.01913]
[47] Grilli F *et al* 2009 *IEEE Trans. Appl. Supercond.* 19 2859-62
[48] Brambilla R *et al* 2009 *Supercond. Sci. Technol.* 22 075018
[49] Morandi A *et al* 2018 *IEEE Trans. Appl. Supercond.* 28 3601310
[50] Fabbri M *et al* 2009 *IEEE Trans. Magn.* 45 192-200
[51] Perini E, Giunchi G, Geri M and Morandi A 2009 *IEEE Trans. Appl. Supercond.* 19 2124-8
[52] Morandi A 2012 *Supercond. Sci. Technol.* 25 104003
[53] Hong Z and Coombs T A 2010 *J. Supercond. Nov. Magn.* 23 1551-62
[54] Zermeno V M R *et al* 2013 *J. Appl. Phys.* 114 173901
[55] Berrospe-Juarez E, Zermeno V M R, Trillaud F and Grilli F 2019 *Supercond. Sci. Technol.* 32 065003
[56] Shen B, Grilli F and Coombs T 2020 *Supercond. Sci. Technol.* 33 033002
[57] Shen B, Grilli F and Coombs T 2020 *IEEE Access* 8 100403-14
[58] HTS Modelling Workgroup: Shared models [http://www.htsmodelling.com/?page_id=748]

**Corrigendum: A new benchmark problem for electromagnetic modelling of superconductors: the high-*T*c superconducting dynamo (2020 *Supercond. Sci. Technol.* 33 105009)**

Mark Ainslie[1], Francesco Grilli[2], Loïc Quéval[3], Enric Pardo[4], Fernando Perez-Mendez[1], Ratu Mataira[5], Antonio Morandi[6], Asef Ghabeli[4], Chris Bumby[5] and Roberto Brambilla[7]

[1]Department of Engineering, University of Cambridge, United Kingdom
[2]Institute for Technical Physics, Karlsruhe Institute of Technology, Karlsruhe, Germany
[3]Group of Electrical Engineering Paris (GeePs), CentraleSup´elec, University of Paris-Saclay, France
[4]Institute of Electrical Engineering, Slovak Academy of Sciences, Slovakia
[5]Robinson Research Institute, Victoria University of Wellington, New Zealand
[6]University of Bologna, Bologna, Italy
[7]Retired; Ricerca sul Sistema Elettrico, Milano, Italy (formerly)

E-mail: mark.ainslie@eng.cam.ac.uk

Since publication, we have realized that the results from the Minimum Electro Magnetic Entropy (MEMEP) method (see section 3.3 in the original paper [1]) used a different voltage definition, causing a slight discrepancy with the results obtained from the other methods in figures 3 and 4. Instead of $V_{eq}(t) = -LE_{ave}(t)$ of equation (5) in [1], we used $-\Delta V(t)$ as defined by equation (22) in [2]:

$$\Delta V \approx L \cdot [\partial_t A_{av,J} + E_{ave}(J)].$$

Although both definitions result in the same DC voltage, the instantaneous signal differs, as shown in Figure 1 in this corrigendum. Indeed, the used $\Delta V$ in [1] adds an extra contribution, $\partial_t A_{av,J}$, the vector potential due to the superconducting current.

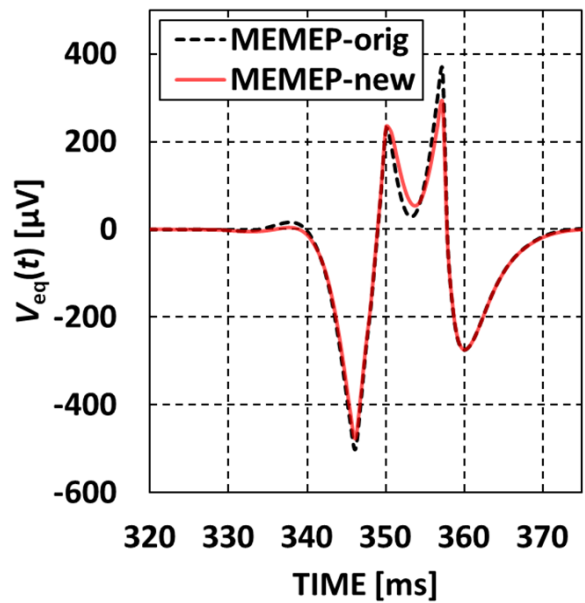

**Figure 1 (Figure 3(d) in the original article [1]):** The new results from MEMEP (red solid line), using the voltage definition for *V*eq given by equation (5) in [1], present better agreement with the other numerical methods than the original results (black dash line), calculated using $-\Delta V$ as described above.

As a result, the new curve has much better qualitative and quantitative agreement with the waveforms calculated by the other methods shown in Figure 3 of the original article [1].

In the original article [1], we also used the $-\Delta V(t)$ definition above instead of $V_{eq}(t)$ to calculate the cumulative time-averaged equivalent voltage, defined from equation (6) in [1] as

$$V_{cumul}(t) = \frac{1}{t}\int_0^t V_{eq}(t)\ dt.$$

The new results, using $V_{eq}(t)$, present only small changes compared to the original calculations, as shown in Figure 2 in this corrigendum.

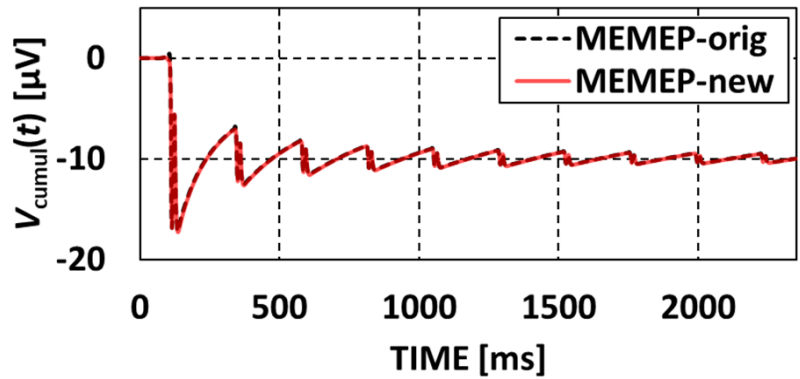

**Figure 2 (Figure 4(d) in the original article [1]):** After using $V_{eq}(t)$ instead of $-\Delta V(t)$ to calculate the cumulative time-averaged equivalent voltage, $V_{cumul}(t)$, the new MEMEP results slightly differ from the original ones.

This curve also has very good qualitative and quantitative agreement with the waveforms calculated by the other methods, as shown in Figure 4 in the original article [1].

The data related to the new calculations for this corrigendum are available at the University of Cambridge data repository (https://doi.org/10.17863/CAM.60437).

Acknowledgements

MA would like to acknowledge financial support from an Engineering and Physical Sciences Research Council (EPSRC) Early Career Fellowship EP/P020313/1. CB, RM and MA were supported in part by NZ Royal Society Marsden Grant No. MFP-VUW1806. FP would like to acknowledge the support of the Consejo Nacional de Ciencia y Tecnologia and Secretaria de Energia de Mexico, Reference No. 541016/439167.

References

[1] Ainslie MD *et al* 2020 *Supercond. Sci. Technol.* 33 105009

[2] Ghabeli A and Pardo E 2020 *Supercond. Sci. Technol.* 33 035008